\documentstyle[a4,11pt,amsmath,graphics,epsfig,citesort]{article}

\newcommand{\eV}{\mbox{\rm eV}}

\newcommand{\MeV}{\mbox{\rm MeV}}

\def\lsim{\raise0.3ex\hbox{$\;<$\kern-0.75em\raise-1.1ex\hbox{$\sim\;$}}}
\def\gsim{\raise0.3ex\hbox{$\;>$\kern-0.75em\raise-1.1ex\hbox{$\sim\;$}}}

\def\npbps#1#2#3{  { Nucl. Phys. }(Proc. Suppl.){\bf B #1} (19#2) #3}
\def\plb#1#2#3{    { Phys. Lett. }{\bf B #1} (19#2) #3}

\begin{document}
\begin{flushright}
{\small
IFUM-828/FT\\
FTUAM-02-514
}
\end{flushright}
\begin{center}
{ \bf
 AFTER SNO AND BEFORE KAMLAND: PRESENT AND FUTURE OF SOLAR AND REACTOR 
NEUTRINO PHYSICS }\\[0.2cm]

{\large P.~Aliani$^{a\star}$, 
V.~Antonelli$^{a\star}$,
R. Ferrari$^{a\star}$,
M.~Picariello$^{a\star}$, 
E.~Torrente-Lujan$^{b\star}$
\\[2mm]
$^a$ {\small\sl Dip. di Fisica, Univ. di Milano},
{\small\sl and INFN,  Via Celoria 16, Milano, Italy}\\
$^b$ {\small\sl Dept. Fisica Teorica C-XI, 
Univ. Autonoma de Madrid, 28049 Madrid, Spain,}\\
}

\end{center}

\vspace{0.2cm}
\begin{abstract}
We present a short review of the existing 
evidence in favor of neutrino mass
and neutrino oscillations which come from different kinds of experiments.
We focus our attention in particular on solar neutrinos, presenting 
a review of some recent  analysis of all available neutrino oscillation evidence in Solar experiments 
 including the recent $SNO\  CC$ and $NC$ data.
We present  in detail the power of the 
reactor experiment KamLAND for discriminating existing 
solutions to the SNP and giving accurate information on 
neutrino masses and mixing angles.

\vspace{5.8cm}

{\large \it \noindent 
Expanded version of the contribution to 
appear in the Proceedings of 
''Third Tropical Workshop on Particle Physics and Cosmology: 
Neutrinos, Branes and Cosmology (Puerto Rico, August 2002)''} 
\end{abstract}

\vfill
{\small \noindent 
$\star$ email: paul@lcm.mi.infn.it, vito.antonelli@mi.infn.it,
 marco.picariello@mi.infn.it, torrente@cern.ch}

\vphantom{\vspace{3cm}}
\newpage

\section{Introduction}

Neutrino has a long and fascinating history that started more than seventy
years ago, when Pauli postulated its existence~\cite{Pauli} to explain the 
data about the high energy part of the $\beta$ decay spectrum. Since then 
the main issue was to discriminate wether neutrino is a massive
or massless particle and eventually to find the value of its mass.

In this work we briefly review the main steps that have been done towards the 
solution of the neutrino mass puzzle and the different kind of experiments that 
have been studying it. Then we remind and critically discuss the essential 
data that have been obtained in this last exciting period and the ones that 
should be presented very soon, focusing in particular our attention on solar and 
reactor neutrino physics. 
We also present the salient aspects of the phenomenological analysis of these 
data that have been developed by our group, with the aim to understand at 
which level  of accuracy we can determine the values of the mixing parameters 
and how we can use the results of the forthcoming experiments to improve this 
accuracy.

The first studies of neutrino mass, performed in the 40th~\cite{primamassa}, were based on the so called Fermi-Perrin method of observation of the $\beta$ spectrum near the end point and they gave the limit $m_{\nu} \leq 500 \, {\rm eV}$.
This limit has been significantly lowered in the following decades and by now 
we know, for instance, that $ m (\nu_e) \leq 2.2{\rm eV}$.
About ten years later than the first measurement of neutrino mass, Goldhaber 
in '58 measured its helicity~\cite{Goldhaber} and proved that the neutrinos pruduced in $\beta$ decays behave like left-handed particles.
 
An essential idea for the understanding of neutrino properties was the 
oscillation hypothesis introduced by Pontecorvo in '57~\cite{Pontecorvo}. 
He conjectured that the neutrino states produced in weak interactions could be
 superpositions of two different states of Majorana neutrinos,
having different values of the mass; therefore there could be 
an oscillation from one neutrino state into another, in a way very similar to
what happens in a neutral kaon system.
This revolutionary idea (implying the existence of at least one mass eigenvalue different from zero for neutrinos) has been further developed in a series of 
works~\cite{otheroscillation} and it has been confirmed by many experimental 
evidences in the following years, as we will see in the rest of this paper. 
Therefore, by now, we are almost sure that neutrinos are massive and 
oscillating particles.      

The experimental hints in favor of a non zero neutrino mass cannot be 
accomodated in the usual ``minimal'' version of the Standard Model of 
electroweak interactions, in which neutrino is simply a left-handed particle 
and it is impossible to build a renormalizable neutrino mass term. On the 
other hand, the existence of such a mass term is predicted  in most of the 
theories beyond the Standard Model, like, for instance, supersymmetric and 
unified theories.
Even if the main issue whether neutrino is massive or not seems to be solved, 
there are still essential theoretical questions to answer. We don't know yet, 
for instance, whether neutrino is a Dirac or Majorana particle, neither we have
 an unique explanation of the reason for which it is so lighter than all the 
other known massive particles. The mechanism usually invoked to explain 
neutrino lightness is the so called see-saw mechanism~\cite{seesaw}, but 
there are also other possible alternative explanations. 
We don't know yet the absolute value of neutrino masses (experiments up to now can only put upper limits and give information on the mass differences)
and which is the exact structure of neutrino mass spectrum. 
We are not able, for instance, to decide whether we are in presence of ``normal'' or ``inverted hierarchy'' or if the masses are ``quasi degenerate''~\cite{massmodels}.

For all these reasons, we can say that in the following years neutrino physics
 will continue to be an ideal playground for testing different theories beyond
 the Standard Model, like, for instance, the supersymmetric and unified 
theories~\cite{susyunified} or the ones predicting the existence of large 
extradimensions~\cite{extradimensions}. This last class of models, which 
become very popular in the last years, have important consequences also for neutrino phenomenology. For instance, they predict the existence of an infinite tower of sterile neutrinos and they can give rise to mechanism explaining the lightness of neutrino mass, similarly to the see-saw, but without the need of 
having a very high mass scale.
However the recent experimental results, mainly from cosmology, put very stringent constraints on the large extradimension theories and ruled out the simpler versions of these models.~\footnote{About the recent status of the models predicting large extradimension and the relative experimental constraints see, for instance, the talk of J. Liu at this conference}.

The experiments aiming to measure neutrino mass or at least to find evidences
that it is different from zero can be classified in three big categories:
\begin{itemize}
\item{the direct kinematical searches;}
\item{the searches of neutrinoless double $\beta$ decay;}
\item{the experiments looking for neutrino oscillations as a proof that neutrinos are massive.}
\end{itemize}

The direct kinematical searches are performed by looking at the high energy 
part of the $^3 H \, \beta$-decay spectrum.
The present limits on the values of $\nu_{\tau}$ and $\nu_{\mu}$ 
masses are~\cite{nutau,numu}: 
\begin{eqnarray}
m (\nu_{\tau}) & < & 18.2 \quad {\rm MeV}\nonumber\\  
m (\nu_{\mu})  & < & 190 \quad {\rm KeV} \, . 
\end{eqnarray}
The best limits for the mass of the electron neutrino, instead, have been obtained from the Mainz and the Troitsk~\cite{Bonn:tw,Lobashev:uu} experiments that 
have found $m (\nu_{e})  < 2.2 \quad {\rm eV} $. 
In future many experiments will try to lower this limit. In particular there 
is a great expectation for KATRIN
(the Karlsruhe Tritium Neutrino experiment)~\cite{KATRIN}, that should 
start the data taking in 2007 and improve the sensitivity down to 
$0.35 \, {\rm eV}$.  

The search for neutrinoless double $\beta$ decays is very important because 
the observation of these decays (violating by two units the lepton number 
conservation) would imply, under the assumption of CPT invariance, the 
Majorana nature of neutrino~\cite{Bilenky:ty,Feruglio:2002af}.
The most stringent limits available at the moment for the effective Majorana 
mass come from the Heidelberg-Moskow 
collaboration~\cite{Klapdor-Kleingrothaus:yx} 
$ \langle m_{\nu} \rangle < 0.35 \, {\rm eV}$ and 
from IGEX (International Germanium Experiment)~\cite{Aalseth:2002rf} 
$\langle m_{\nu} \rangle < 0.33-1.35 \, {\rm eV}$. 

The effective Majorana mass is given by:
\begin{eqnarray}
\left|\langle m \rangle| = |m_1 |U_{e1}|^2 + m_2 |U_{e2}|^2 e^{i \alpha_{21}} + m_3 |U_{e3}|^2 e^{i \alpha_{31}}\right|,
\label{massmajorana}
\end{eqnarray}
where $m_i$ indicate the values of the mass eigenstates, the $U_{ei}$ are elements of the Pontecorvo-Maki-Nakagawa-Sakata mixing matrix and $\alpha_{21}, \alpha_{31}$ are the two additional CP violating Majorana phases.
The masses and the elements of the mixing matrix entering equation~(\ref{massmajorana}) can be expressed in terms of the mass differences and mixing angles that can be recovered from the experiments on solar, atmospheric and reactor neutrinos. Thus, given these experimental inputs, the value of $|\langle m \rangle| $ depends on the ones of $m_1$ and of the Majorana phases and on the general 
structure of neutrino mass spectrum (i.e. normal or inverted hierarchy or quasi degenerate case). In particular one can get interesting lower bounds in the cases of inverted hierarchy and quasi degenerate mass spectrum and a stringent upper bound~\cite{limitem_e} (smaller than $10^{-2} {\rm eV}$) in the case of 
normal hierarchy.
The NEMO3~\cite{nemo3} experiment and the criogenic detector 
CUORICINO~\cite{Pirro:fi} are expected to reach sensitivity of the order of $10^{-1} {\rm eV}$ and even an order of magnitude better is planned for the next generation experiments~\cite{Pirro:fi,altridoppiobeta} CUORE, GENIUS, EXO, MAJORANA and MOON.
Therefore, there is the hope that the results of these future experiments, 
complemented with the information on the absolute value of the masses that 
could come from KATRIN and from cosmology, could be used to recover hints 
of CP violation in the leptonic sector or to discriminate between the different
possible structures of the mass spectrum.
For a more detailed analysis of the present situation and the future perspectives of neutrinoless double $\beta$ decays we refer the interested reader to the
work~\cite{Petcovultimo} and the references contained in it.

A first group of experiments looking for oscillation signals use neutrino 
fluxes produced at accelerators and nuclear reactors. 
They are usually distincted in long- and short baseline, according to the 
distance existing between the neutrino production point and the detector.
Many short baseline accelerator experiments, like NOMAD~\cite{NOMAD} and 
CHORUS~\cite{CHORUS} at Cern, didn't find any signal of oscillation
and, consequently, they gave important constraints on the possible values of 
the mixing parameters. 
Between the reactor experiments, it is worthwhile to remember 
the results of CHOOZ~\cite{CHOOZ} and Palo Verde~\cite{Boehm:2001ik}. 
At CHOOZ a beam of reactor $\bar{\nu}_e$ was sent to a detector located 
about 1 Km far away and it was detected through the reaction 
$\bar{\nu}_e +p \to e^+ + n$.  No evidence of oscillation was found 
and these results can be used to exclude a significant part of the 
mixing parameters plane. In particular they tell us that $\Delta m^2$ must be 
smaller than $10^{-3} eV^2$, unless the values of the mixing angle are very 
small. The CHOOZ data can be used also to fix an  upper limit for the quantity $sin^2 (\theta)$, where $sin^2 \theta$ coincides with $|U_{e3}|^2$ in the case 
of normal hierarchy and $sin^2 \theta \equiv |U_{e1}|^2$ for neutrinos with inverted hierarchy mass spectrum.

The opposite situation took place in the case of LSND~\cite{Sung:ps}, a 
short baseline accelerator experiment performed with a neutrino beam produced 
at the Los Alamos meson physics facility (LAMPF).      
This experiment found evidences for two kind of oscillation signals: 
that is an
excess of $\bar{\nu}_e$ in the beam of $\bar{\nu}_{\mu}$ produced by 
the decay at rest of the $\mu^+$ (obtained as secondary products of the 
proton accelerator beam), 
and a signal of oscillations into $\nu_e$  of the $\nu_{\mu}$ 
produced by the $\pi^+$ decay in flight.
The LSND result, if confirmed, would be a clear indication of 
oscillation with very high values of the mass difference, up to $\Delta m^2 
\simeq 1 {\rm eV}^2$.
To reconciliate this result with the ones coming from solar and atmospheric 
neutrinos (that we are going to present), one would have to postulate the 
existence of at least one sterile neutrino in addition to the usual three 
active ones.
However, up to now there have been no independent confirmations of the LSND 
results.
The KARMEN experiment~\cite{Wolf:2001gu}, 
performed at the Rutherford 
Laboratories, explored a significant part of the mixing parameter space 
proposed by LSND and it didn't find any signal of oscillation 
\footnote{About the compatibility of LSND and Karmen results see 
also~\cite{LSNDconKARMEN}}. 
A new experiment MiniBoone~\cite{Hawker:pt,Stefanski:2001rk} just started running and it shuold soon produce data. It has been projected in such a way to 
test definitely the validity of LSND results. 
     
A new generation of very long baseline experiments became available in the 
last years. The forerunner of them is K2K~\cite{Yanagisawa:xz,K2K}, that uses 
a neutrino beam produced at the Japan kaon facility KEK and 
detected at the Kamioka site. Up to now K2K has detected 56 events instead 
of the expected value in absence of oscillations, that is $80^{+7.3}_{-8}$. 
This is a confirmation of neutrino oscillations (the null oscillation 
probability is less than $1 \%$). Moreover the best fit point~\cite{K2K} 
values for the mass difference  and the mixing angle 
($\Delta m^2 = 2.8 \times 10^{-3} {\rm eV}^2$ and $\sin^2 2\theta = 1$) are in good 
agreement with the results of atmospheric neutrino experiments.
Two similar projects have been already approved and will become available 
in the near future: one of them is a   
neutrino beam from CERN to the Gran Sasso Labs~\cite{CernGSasso}  
and the other one is in the States~\cite{longUSA} (from FNAL to Soudan). 
The long baseline accelerator experiments will probably give
an important confirmation of the oscillation evidences coming up to now from 
the study of solar and atmospheric neutrinos. They are also expected to 
find in an unambiguous way indications of oscillation from appearance signals. 
In the long baseline experiment one has in addition the opportunity of 
choosing the specific characteristic of the beam; hence they can be used to 
perform precision measurements. For instance they should be useful to study 
the value of the mixing angle $\theta_{13}$, relevant for eventual CP 
violation. 
The present limit on the measurement of this angle coming from CHOOZ 
($\theta_{13} \leq 9$ degrees), could be lowered to the level of about 
$5$ degrees at ICARUS, one of the two experiments that will use the 
CERN-Gran Sasso beam.

Important result should very soon come also from the long baseline reactor 
experiment kamLAND~\cite{kamLAND}, that could in principle give a definite 
solution to the solar neutrino problem, as we are going to see in the following.   

The two main categories of experiments looking for oscillation signals are 
the ones that study the atmospheric and the solar neutrinos.

The flux and the angular distribution of electronic and muonic neutrinos 
produced in the atmosphere, in the processes generated by cosmic rays, is known 
with quite a good accuracy~\cite{Battistoni:2002ew}, taking into account the 
number and the properties of cosmic rays and the important geomagnetic effects.
This input information are essential, because most of the atmospheric neutrino 
experiments measure  the value of the double ratio 
$R= \frac{(\mu/e)_{data}}{(\mu/e)_{MC}}$. The numerator and denominator 
 are, respectively, the experimental and the Monte Carlo computed values of 
the ratio between the events generated by muonic neutrinos (and antineutrinos)
and the ones generated by electronic neutrinos (antineutrinos).

At present there are different experiments running on atmospheric 
neutrinos and most of them are essentially water Cherenkov detectors 
(like Kamiokande~\cite{Hatakeyama:1998ea,Oyama:bk,Fukuda:1994mc,Hirata:1992ku}, 
Super-Kamiokande
~\cite{Fukuda:1998mi,Kajita:zv,Toshito:2001dk,Fukuda:2000np}, IMB~\cite{Becker-Szendy:vr}) or iron plate calorimeters 
(like Soudan II~\cite{Petyt:rn} and in the past years Frejus~\cite{Daum:bf} 
and Nusex~\cite{Aglietta:1988be}). 
Clear evidences of oscillation have been found at Kamiokande, 
Super-Kamiokande (SK), IMB and Soudan II and also at the 
MACRO~\cite{Giorgini:2002it} experiment at Gran Sasso. 

Here we recall the results of SK, which are based on 
a very high statystic and are, in any case, in good agreement with the other 
atmospheric experiments which found oscillation signals. The SK results are $R = 0.638 \pm 0.016 \pm 0.050$ 
for the Sub-GeV events and $R = 0.658^{+0.030}_{-0.028} \pm 0.078$ for the 
Multi-GeV events.
Another interesting observable is the up-down asymmetry between the up 
going events, in which the neutrino crossed the Earth before interacting 
in the detector, and the down going ones: 
$A_{e,\mu}= (\frac{U-D}{U+D})$.   
The experimental value of this quantity is consistent with zero for the 
electronic neutrinos, while for the muonic ones the day-night asymmetry
for high values of the momenta is a decreasing negative value. 
This result is a clear indication of a reduction of the flux of 
muonic neutrinos and antineurinos that arrive to the detector after crossing
the Earth. 
The most natural explanation of this phenomenon is the possibility that the 
muonic neutrinos oscillate into other flavors and the oscillation probability
is greatly enhanced by the interaction with matter.

The last group of experiments is that of the experiments observing the 
neutrinos coming from the Sun. We will discuss them in detail in the rest of 
the paper. 


\subsection{History of the solar neutrino problem}

The first experiment on solar neutrinos, Homestake~\cite{Homestake}, started at the end 
of the $`60s$ using the inverse $\beta$ decay on chlorine 
$^{37}Cl + \nu_e \to  ^{37}Ar + e^-$. The threshold energy was 
$E_{thr} \simeq 0.81 {\rm \MeV}$, hence it was sensitive to the pep, $^7 Be$, 
$^8 B$ and hep components of the solar neutrino flux.
The results were really surprising, because Homestake found a deficit of the solar
neutrino flux of more than $60 \%$ that predicted by the Solar 
Standard Model (SSM). The updated value of the ratio $R$, between the 
experimental results and the SSM prediction~\cite{BPB2000}, for the chlorine  
experiment is $R=0.34 \pm 0.03$.
This result raised fundamental questions: what happens to solar $\nu$ on their
 way to earth? Eventually, could the SSM be wrong? 

The Homestake indication was confirmed by similar experiments, SAGE~\cite{SAGE}
in Russia and GALLEX~\cite{GALLEX} and later on GNO~\cite{GNO} at the INFN 
Gran Sasso Labs, which used gallium instead of chlorine.
The energy threshold is lower in the gallium experiments 
($E_{thr} \simeq 233 \, keV$) making them also sensitive to pp
neutrinos, which are the main component of the solar neutrino flux.  
The updated gallium results are~\cite{SAGE,GALLEX,GNO}:
\begin{eqnarray}
R & = & 0.60 \pm 0.05 \quad (SAGE) \nonumber\\  
R & = & 0.58 \pm 0.05 \quad (GALLEX-GNO) \, . 
\end{eqnarray}
This confirmation of Homestake results gave a strong support to the neutrino 
oscillation hypothesis and caused an increase of the interest for this problem.
It could be a signal of {\it new physics}.

An essential improvement in the knowledge of solar neutrinos came with the 
advent of the water Cherenkov experiments, Kamiokande~\cite{Kamiokande} and
Super-Kamiokande (SK)~\cite{SKsolar,Smy:2002hr}, that 
looked at the elastic scattering $\nu_e + e^- \to \nu_e + e^-$ and 
confirmed the existence of the ``solar neutrino problem'' with a very high 
statistic. In this experiments it was possible to know the direction of the 
incoming neutrino (by looking at the outgoing direction of the recoil electron)
and also to study the energy and angular spectrum and the day night 
asymmetries.  
The energy threshold for these experiments was quite high (5 \MeV for SK) 
and therefore they were sensitive only to the high energy component of the neutrino flux, that is $^8 B$ and hep neutrinos. Their results confirmed the existence of a deficit in the electron neutrinos reaching the detector. 
The SK result for the energy spectrum and the small values of the day-night 
asymmetries were also very important to put strong constraints on the possible
 values of the mixing parameters.  

After the publication of SK data it was clear that there was a deficit 
of solar electron neutrinos reaching the Earth, with respect to the flux predicted by SSM. The oscillation hypothesis was considered the most 
plausible explanation of this phenomenon, but there were still different 
regions allowed by the experiments in the mixing parameter plane, as we will 
see in detail. 

\subsection{The post SNO situation}

The real breakthrough was due to the SNO experiment that published its first 
data in 2001~\cite{SNOCC} ( see also \cite{Bahcall:2001zu}). 
SNO is a deuterium Cherenkov detector designed to 
look simultaneously at three different reactions:
\begin{equation}
\begin{array}{lllll}
\nu_e + d & \to & e^- + p +p &  \text{{\sl (Charged Current)};}\\
\nu_x + d & \to & e^- + n +p & \text{{\sl (Neutral Current)};}\\  
\nu_x + e^- & \to & \nu_x +e^-  & \text{{\sl (Elastic Scattering)}.} 
\end{array}
\end{equation}
The first reaction (CC) receives contribution only from the electron neutrino,
while the others (NC and ES) are sensitive to all neutrino flavors.
This experiment gives the first direct model independent measurement of the 
total solar neutrino flux reaching the Earth (through the NC observation) and 
at the same time, comparing this flux with the one of $\nu_e$ recovered from 
CC, it offers a strong evidence of the oscillation of $\nu_e$ into other 
active neutrinos.

During its first phase of working~\cite{SNOCC} SNO observed the charged current
and elastic scattering events, with an energy threshold for electron
detection of 6.75 \MeV.
The $\nu_e$ flux measured from CC, after 241 days of running, was: 
$\Phi_{\nu_e}^{CC} = 1.75 \pm 0.07 (stat.)^{+0.12}_{-0.11} (syst.) \times 10^6 
cm^{-2} s^{-1}$. The ratio between this value and the SSM prediction was 
$R=0.35 \pm 0.03$. 
In the SNO experiment the neutrino flux can be recovered also from the elastic
scattering, using the relation \footnote{the contribution of $\nu_{\mu}$ 
and $\nu_{\tau}$ to the elastic scattering cross section is only through 
neutral current, hence it is about 1/6 of the contribution of $\nu_e$ that can 
interact also through charged current}:
\begin{equation}
\Phi_{\nu}^{ES} = \Phi_{\nu_e}^{ES} +0.154 \sum_{i=\mu,\tau} 
\Phi_{\nu_i}^{ES}.
\end{equation} 
The value of the total neutrino flux recovered from the elastic scattering at 
SNO and also, with a better statistics, at SK doesn't agree with the $\nu_e$
flux obtained from SNO CC. The comparison of the two results gives:
\begin{equation}
\sum_{i=\mu,\tau} \Phi_{\nu_i}^{ES} = 3.69 \pm 1.13 \times 10^6 cm^{-2} 
s^{-1}. 
\end{equation}
This result was the first evidence (at $3 \sigma$ level) of the 
presence of muonic and tauonic neutrinos in a electronic neutrino beam 
reaching the Earth from the Sun. Therefore it was, up to the present SNO
data on NC, the most robust evidence of $\nu_e$ oscillation into other active 
neutrinos.
It is also remarkable that the sum of the $\nu_e$ and $\nu_{\mu,\tau}$ 
fluxes give a value in good agreement with the SSM prediction. Consequently 
the results of SNO phase I also strongly disfavored the 
hypothesis of pure oscillation into sterile neutrinos.

Observation of neutral current neutrino interactions on deuterium in 
 the SNO experiment has been recently presented
~\cite{Ahmad:2002jz,Ahmad:2002ka}.
Using the NC, ES and CC reactions and assuming the ${}^8 B$ neutrino 
 shape predicted by the SSM, the electron and active non-electron neutrino component
 of the solar flux at high energies ($\gsim $5 MeV) are obtained.
 The non-electron component is found to be $\sim 5\sigma $ greater than 
zero, the standard prediction, thus providing the strongest evidence so 
 far for flavour oscillation in the neutral lepton sector:
the agreement  of the total flux, provided by the NC measurement
 with the expectations 
 implies as a by-product the confirmation of the validity of the
 SSM~\cite{turck,bpb2001,bp95}.

The  results presented recently by $SNO$ on Solar neutrinos 
\cite{sno2001} 
 confirm and are consistent with previous evidence from SK 
and the rest  solar neutrino 
experiments~\cite{homestake,Fukuda:1999rq,Fukuda:1998fd,Fukuda:1998fd,sage1999}.
The CC, ES and NC global and day and night 
fluxes  presented in Refs.~\cite{Ahmad:2002jz,Ahmad:2002ka}. 
 are derived under 
 the assumption that the ${}^8 B$ spectral shape is 
not distorted from the SSM prediction. 
With this assumption the SNO collaboration  checks 
the hypothesis of non-oscillation, or zero $\phi_{\mu+\tau}$ flux.

Different quantities can be defined 
\cite{Aliani:2002ma}
in order to make  the evidence for 
disappearance and  appearance of the neutrino flavours explicit. 
Letting alone the SNO data, from the three fluxes measured by SNO is possible to 
define two useful ratios, deviations of these ratios with respect to  their 
standard value are powerful tests for occurrence of new physics.
Here we compute the
values for $\phi_{CC}/\phi_{ES}$ and 
$\phi_{CC}/\phi_{NC}$  being specially careful with the treatment of the 
correlations on the uncertitudes, the inclusion or not of these correlations
can affect significantly the results for these 
ratios (see table II in Ref.~\cite{Ahmad:2002jz}.
 for a complete
 list of sistematical errors).
For the first ratio, from the value from SNO rates
~\cite{Ahmad:2002jz,Ahmad:2002ka}.
 we obtain
$$ \frac{\phi_{CC}}{\phi_{ES}}=0.73^{+0.10}_{-0.07}, $$
 a value which is  $\sim$ 2.7 $\sigma$ away from the  no-oscillation expectation value of one.
The ratio of CC and NC fluxes gives the fraction of electron neutrinos remaining 
in the solar neutrino beam, our value is:
$$\frac{\phi_{CC}}{\phi_{NC}}=0.34^{+0.05}_{-0.04},$$
this value is nominally many standard deviations ($\sim 13 \sigma$) away  from the 
 standard model case \cite{Bahcall:1996bw}.  

Finally, if in addition to SNO data  we consider the 
 flux predicted by the solar standard mode one can define, 
following   Ref.\cite{Barger:2001zs},  the 
quantity $\sin^2\alpha$, the fraction of 
''oscillation neutrinos which oscillated into active ones'', again using 
 the SNO data and fully applying systematic correlations 
(see table 2 in Ref.~\cite{Ahmad:2002jz,Ahmad:2002ka}), we find the 
following result:
$$
\sin^2\alpha=\frac{\phi_{NC}-\phi_{CC}}{\phi_{SSM}-\phi_{CC}}=0.92^{+0.39}_{-0.20}.
$$
The SSM flux is taken as the ${}^8\rm B$ flux predicted in Ref.\cite{bpb2001}.
Note that, although consistent with it, this result differs significantly from the number obtained in 
Ref.\cite{Barger:2001zs}, this is due to the introduction of 
 systematic correlations in our calculation. 
The central value is clearly below one 
(only-active oscillations):
the fraction of sterile neutrinos 
is $\cos^2\alpha \lsim 0.28$ 
(1$\sigma$).
Although electron neutrinos are still allowed to oscillate into 
sterile neutrinos the hypothesis of transitions to {\em only} sterile 
 neutrinos is rejected at nearly $5\sigma$, this significance  would be reduced  if we consider applying
 a 1-sided analysis to avoid non-physical values.

\section{Global analysis of the solar neutrino data}

Given all the experimental data that we have just reported, one can 
say that there is really strong evidence that neutrinos are massive 
and oscillating particles. Nevertheless, details of the 
mass pattern still have a long way before  being clarified.
With this aim in mind, we developed a global analysis of all the available data  
on solar neutrinos, also including the CHOOZ constraints.  
Our first purpose was that of determining the regions in the mixing parameter 
plane that are still compatible with the experiments.
In addition to this, we wanted to understand how the forthcoming experiments
(in particular Borexino and KamLAND) could improve our knowledge of neutrino 
mass properties.

We assumed neutrino oscillation as a working hypothesis and considered 
bidimensional models. 
For details of our analysis we refer the interested reader 
to~\cite{Aliani:2001zi,Aliani:2002ma}.
Here we just report the most salient aspects of our strategy.
The analysis is based on the numerical calculation of the expected 
event rate for every solar neutrino experiment as a function of the mixing 
parameters and on the comparison between these expected numbers and the 
experimental data.
The statistical analysis is based on the $\chi^2$ method its output being
 contour plots in which one can see which values of the mixing angles
and mass differences are still allowed at a given confidence level.

Our calculation can  essentially be split into two parts. The first one is 
the determination of the neutrino transition amplitude, i.e. the 
probability for an electronic neutrino produced in the Sun to change its 
flavor before reaching the detector.  
The other ingredient is the calculation of the detector response functions, 
that, for a given neutrino energy, depend on the experimental details of the 
specific detector (i.e. efficiency, resolution, etc.) and on the cross 
section for the reaction under examination.

The transition amplitude calculation is separated in three parts, 
corresponding to the neutrino propagation inside the Sun, in the vacuum and in the Earth. For every value of the mixing parameters we compute fully 
numerically the amplitudes in the Sun and in the Earth, while the one corresponding to the vacuum evolution is computed analytically. The three amplitudes are patched together using the evolution operator formalism
\cite{Torrente-Lujan:1998sy}. 

\section{The present situation}

\begin{figure}[tb]
\centering
\psfig{file=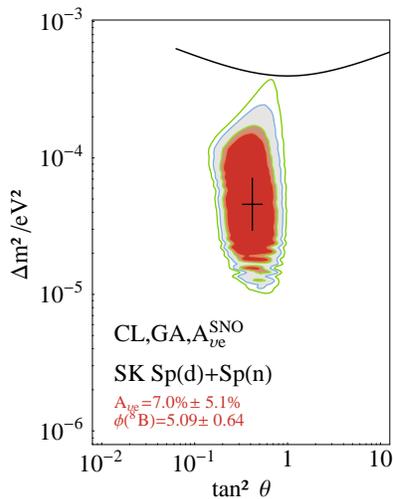,width=6cm} 
\caption{\small
Allowed areas in the two neutrino parameter space.
The point with error bars corresponds to best 
the results from fit to marginal likelihoods.
The coloured areas are the allowed regions at 
90, 95, 99 and 99.7\% CL relative to the absolute minimum.
The region above the upper thick line is excluded by the 
reactor experiments \protect\cite{chooznew}.
}
\label{sno:f1}
\end{figure}

\begin{figure}[p]
\centering
\begin{tabular}{ll}
\psfig{file=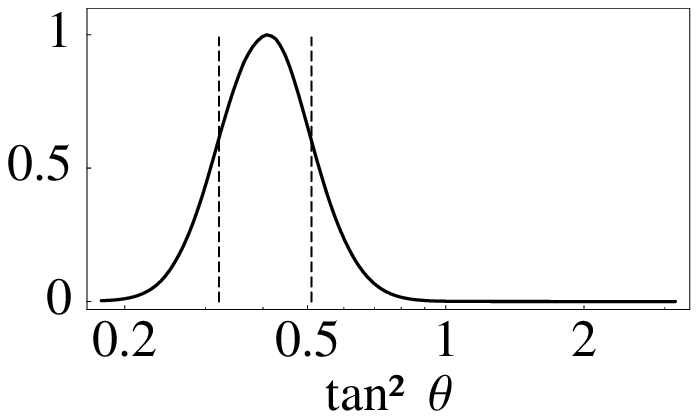,width=8cm,height=5cm} &\hspace{-1.5cm}
\psfig{file=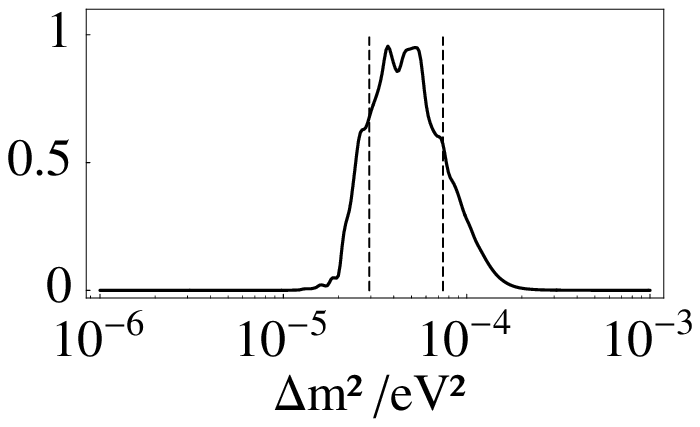,width=8cm,height=5cm} 
\end{tabular}
\caption{\small
Marginalized likelihood distributions for each of the
 oscillation parameters 
$\Delta m^2$ (right), $\tan^2 \theta$ (left).
The curves are in arbitrary units with normalization to the 
 maximum height.
Values for the peak position are obtained by fitting 
two-sided gaussian distrbutions (not showed in the plot).
Dashed lines delimit $\pm 1\sigma$ error regions around the maximum.}
\label{f2}
\end{figure}

We included in our analysis 
\cite{Aliani:2002ma}
the total rates of the chlorine and gallium 
experiments, together with the different energy bins of SuperKamiokande and  
with the most recent results of SNO~\cite{Ahmad:2002jz,Ahmad:2002ka}.
The results are shown in Fig.\ref{sno:f1} where we have generated acceptance 
contours in $\Delta m^2$ and $\tan^2\theta$.
In  Table~(\ref{table2})
 we present the best fit parameters or local minima 
 obtained from the minimisation of the $\chi^2$ function.
Also shown are the values of $\chi^2_{\rm min}$ per degree of freedom
 ($\chi^2/n$) and the goodness of fit (g.o.f.) or significance level of each 
 point (definition of SL as in Ref.~\cite{numu}).
In order to obtain concrete values for the individual 
oscillation parameters and 
estimates for their uncertainties, it is preferable
to study the marginalized parameter constraints.
It is justified to convert $\chi^2$ into likelihood
 using the expression ${\cal L}=e^{-\chi^2/2}$, this normalised marginal 
likelihood is plotted
 in Figs.~(\ref{f2}) for each of the oscillation parameters $\Delta m^2$ and $\tan^2\theta$.
For $\tan^2\theta$ we observe that the likelihood function is concentrated in a 
 region $0.2<\tan^2\theta<1$ with a clear maximum at $\tan^2\theta\sim 0.4$.
The situation for $\Delta m^2$ is similar.
Values for the parameters are extracted by fitting  one- or two-sided 
Gaussian distributions to any 
of the peaks (fits not showed in the plots). In the case of the angle distribution the goodness of fit of the 
Gaussian fit is excellent (g.o.f $>99.9\%$) even at far tail distances thus 
justifying the consistency of the procedure. The goodness of Gaussian fit to
the distribution in squared mass, although somewhat smaller, is still  good.
The values for the parameters appear in Table~\ref{table2}. They are 
fully consistent and very similar to the values obtained from simple 
$\chi^2$ minimisation. 

\begin{table}[h]
 \begin{center}
  \scalebox{0.95}{
    \begin{tabular}{lllll}
 Method  & & & & \\
\hline
 & & & & \\
A) Minimum LMA   & $\Delta m^2=5.44\times 10^{-5}\ \eV^2$ & $\tan^2\theta=0.40$    & $\chi_{m}^2$ = 30.8&  g.o.f.: 80\% \\
    & & & & \\
B) From Fit    & $\Delta m^2= 4.5^{+2.7}_{-1.4}\times 10^{-5} \eV^2$ & $\tan^2\theta=0.40^{+0.10}_{-0.08}$   &   &   \\
 & & & & \\
\hline
      \end{tabular}
}
 \caption{\small
Mixing parameters: 
A) Best fit in the LMA region and B) 
from fit to marginal likelihood distributions.}
  \label{table2}
 \end{center}
\end{table}

In summary,
the direct measurement via the $NC$ reaction on deuterium of 
${}^{8}{\rm B}$ neutrinos combined with the
$CC$ results have largely 
confirmed the neutrino oscillation hypothesis.
We have  obtained the allowed area in parameter space 
and individual 
values for $\Delta m^2$ and $\tan^2\theta$ with error estimation
 from the analysis of marginal likelihoods.

\section{Future scenarios: Borexino}

Given this situation, we studied which new information should come in future 
from the Borexino data~\cite{Aliani:2001zi,Aliani:2001ba,Aliani:2002rv}. 
Borexino~\cite{Borexino,Meroni:zj} is a solar neutrino experiment, mainly sensitive to 
the $^7 Be$ component of the neutrino flux, that should start running in very 
next years at the Gran Sasso Labs. In Figure 3 the usual contour plots 
obtained from all the experiments available up to now are superimposed to the 
contour lines corresponding to different hypothetical possible values
 of the total rate at Borexino. Note that, for what concerns the data coming from
the other solar neutrino experiments, this figure refers to the situation how it was 
in 2001, before the publication of the last results from SK and from SNO neutral current. 
As one can see from the picture, Borexino should be able to clarify  
the situation in the case in which the solution  very well is in the small mixing angle region. The situation would be, instead, more complicate in case of LMA or LOW 
solutions. In these two regions, in fact, the ratio between the Borexino 
signal and the SSM prediction in absence of oscillations should be between 
0.6 and 0.7 .
The discrimination power of Borexino increases a lot if we look also at the 
day-night asymmetry, as one can see from Figure 4.
The LOW region is characterized by high values of the asymmetry, that can reach
up to $ 20 \%$, while in the LMA region the day-night asymmetry is much lower.
Hence, by looking simultaneously at the total rate and at the day-night 
asymmetry, Borexino should be able to discriminate between the two solutions 
of the solar neutrino problem that are compatible with the experiments up to 
now, that is the LMA and the LOW solutions.

\begin{figure}
\centering
\includegraphics[scale=0.4]{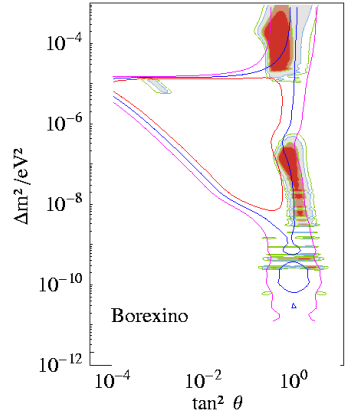}
\caption{\small
 Possible values of the total signal at Borexino. The full lines correspond to a total rate equal respectively to 0.5, 0.6 and 0.7 with respect to the SSM 
prediction in absence of oscillation. These isosignal lines are superimposed   to the contour plots (colored regions) corresponding to the regions allowed 
at different confidence levels from the other experiments.}

\label{Borexinodaynight}
\label{borexinototal}
\end{figure}

\section{Future Experiments: Kamland}

Another very important experiment, already running, that should significantly improve
our knowledge of the mixing parameters relevant for solar 
neutrinos is KamLAND~\cite{kamLAND}. 
In this experiment a flux of low energy $\bar{\nu}_e$ produced by different nuclear reactors is sent to a scintillator detector capable of detecting their interactions with protons.  
Although it not a  solar
neutrino experiment, KamLAND is sensitive to neutrino oscillations with 
mixing parameters in the LMA region, that seems to be the solution of the 
solar neutrino problem preferred by the present data.
Therefore, we can hope that KamLAND will soon be able to determine the exact
values of the mixing parameters with satisfactory accuracy. 
The main limitation of KamLAND is its reduced sensitivity to the extreme 
upper part of the LMA region, that could create problems in the determination 
of $\Delta m_{12}^2$,  as discussed in~\cite{Strumia:2001gi,Petcov:2001sy} and later on in~\cite{HLMA}. 
For a detailed discussion about KamLAND potentiality and discrimination power 
we refer the interested reader 
to~\cite{Aliani:2002ca,Aliani:2002rv}.

In order to study the potentiality of KamLAND for 
resolving the neutrino oscillation parameter space,
we have developed  two kind of analysis. In the first 
case (Analysis A below) we   deal with the KamLAND 
expected global signal. We  asumme that the experiment 
measure a certain global signal 
with given statistical and systematic 
error after some period of data taking 
(1 or 3 yrs) and  perform a complete $\chi^2$ 
statistical analysis including in addition the 
up-date solar evidence. 
In the second case, Analysis B, we  include the full 
KamLAND spectrum information. 
Instead of giving arbitrary values to the different bins, 
we  assume a number of oscillation models characterized 
by their mixing parameters $(\Delta m^2,\theta)$. 
After including the  solar evidence 
we   perform the same $\chi^2$ analysis as before.

\begin{figure}
\centering
\psfig{file=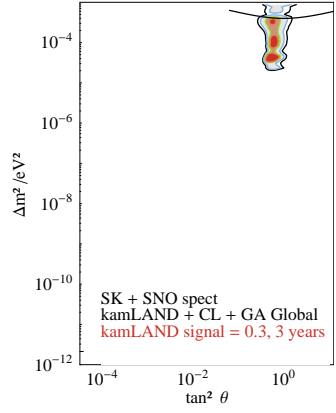,width=5cm}
\caption{
\small Exclusion plots including KamLAND global rates 
(Analysis A), given a
 hypothetical experimental global signal ratio: 
 respectively  $S/S_0= 0.3$.
Statistical and assumed systematics 
($\sim 5\%$) errors are included.  
Three year of KamLAND 
data taking.
The colored areas are allowed   at 
90, 95, 99 and 99.7\% CL relative to the absolute minimum.
The region above the upper thick line is excluded by the 
reactor experiments \protect\cite{chooznew}.
}
\label{kl:f1}
\end{figure}

In Fig.\ref{kl:f1} we graphically show the 
results of this analysis. They represent
exclusion plots including KamLAND global rates, given a
 hypothetical experimental global signal ratio: 
 respectively strong and medium suppression $S/S_0= 0.3,0.6$ 
and no oscillation 
 evidence $S/S_0= 1.0$ for one and three years of KamLAND 
data taking.
As can be seen in the figures, as the 
KamLAND experimental signal decreases, the LMA region is singled 
out. The periodic shape in $\Delta m^{2}$ of the 90 \% C.L. 
(red regions) which becomes apparent in the three-year plot 
 is due to the periodicity of the 
response function: in order to distinguish among these different 
equally-likely solutions, one
 must analyze the energy spectrum, 
Obviously, only if KamLAND sees some oscillation signal 
(i.e. $S_{i}/S_{0} << 1.0$ ) does the LMA region 
become the only solution. If we consider a hypothetical signal 
closer to 1.0 than 0.3, we see that the LOW region survives, 
although it is less favored.

\begin{figure}
\centering
\begin{tabular}{lr}
\psfig{file=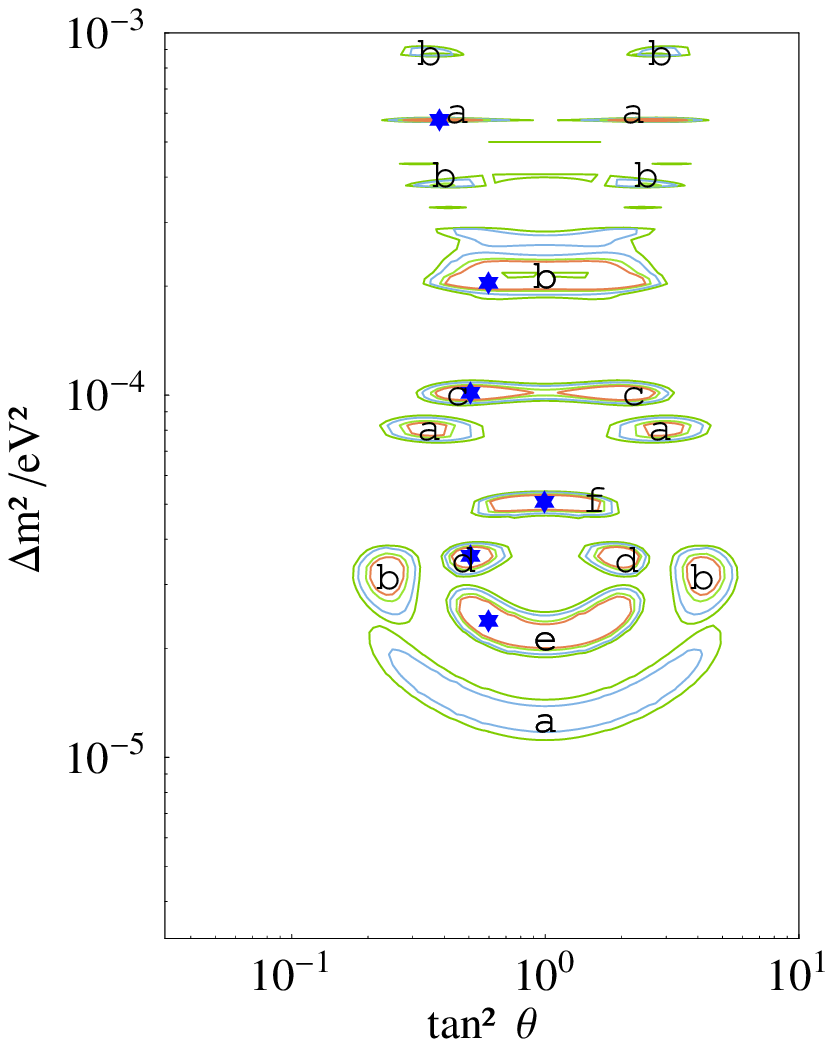,width=5cm}
&\psfig{file=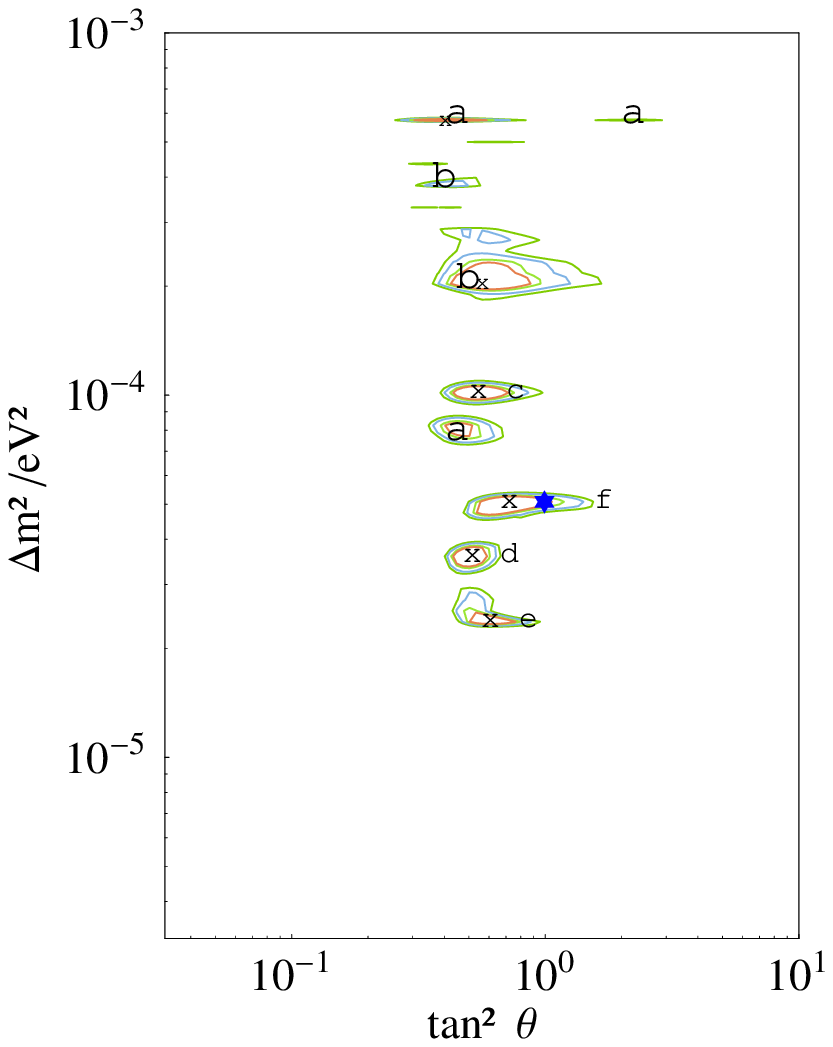,width=5cm}
\end{tabular}
\caption{\small 
Allowed areas in the two neutrino parameter space
after 3 years of data taking in KamLAND
(Analysis B).
Allowed regions belonging to the same point are labeled 
with the corresponding letter, the position of the 
point itself is labeled with a solid star.
The colored lines separate allowed regions at 
90, 95, 99 and 99.7\% CL relative to the absolute minimum.
(Left) Results with the KamLAND spectrum alone.
(Right) KamLAND spectrum plus solar (CL,GA,SK,SNO) evidence.
Crosses are situated in the position of the $\chi^2$ minima.
}
\label{f3b}
\end{figure}

In Figs.(\ref{f3b})  we graphically 
show the results of this analysis for a selection of points
and for three years of data taking restricting ourselves 
to the LMA region of the parameter space where, as we have 
noted before, the KamLAND spectrum information is specially
significative.  
In each plot allowed regions corresponding to different 
starting points are superimposed, every region is 
distinguished with a label. The position of initial points is 
labeled with solid stars.

The first case, study of the KL spectrum alone
 ($\chi^2_{spec,KL}$) is represented by the 
 Fig.(\ref{f3b} left). 
The allowed parameter space corresponding to each 
particular point is formed by a number of,
 highly degenerated, disconnected 
regions symmetric with respect the line $\tan\theta=1$.
These regions can extend very far from the initial point 
specially in terms of $\Delta m^2$ but also in some 
occasions in terms of $\tan^2\theta$.
For example the point ``A'' located at 
$(\Delta m^2=5.7 \times 10^{-4},\tan^2\theta=0.38 )$
gives rise to two sets of thin regions situated  respectively
at $\Delta m^2\sim 10^{-3},10^{-4}$ and a third region 
situated at 
$\Delta m^2\sim 10^{-5}$ 
which practically covers the full range $\tan^2\theta\sim 0.1-10$.
A similar behavior is observed for point ``B''.
Of course this situation is not very  favorable for the 
future phenomenologist trying to extract conclusions 
from the KamLAND data. A much comfortable situation is 
found for points nearer the center of the LMA region.
Note how the regions corresponding to the points 
``D,E'' and specially ``F'' only extend very gently around 
the initial location.

The results of the full analysis are summarized in
 Fig.(\ref{f3b}, right). 
The position of the minima of $\chi^2$, marked in the 
plot with crosses, is practically identical to the position 
of the initial points except in some case where the difference
is not significant anyway.
The general effect of the inclusion of the solar 
evidence in the $\chi^2$ is the breaking of the symmetry in 
$\tan^2 \theta$ as expected and the general reduction 
 of the number of disconnected regions corresponding to 
each point. Note however that the point ``A'' still 
gives rise to a small allowed region situated nearly one
 order 
of magnitude smaller in $\Delta m^2$. The ``B'' region is 
shrinked near its initial location as happens  to the 
rest of points.
The conclusion to be drawed from these plots is that 
KamLAND together with the rest of solar experiments 
will be able to resolve the neutrino mixing parameters with
 a precision of $\delta \log \Delta m^2\sim \pm 0.1$ practically
 everywhere. However, for values of $\Delta m^2> 10^{-4}$
 the problem of the coexistence of multiple regions with 
 similar statistical significance will still be present.

\subsection{KamLAND and solar antineutrinos}

We have also investigated  
\cite{Aliani:2002pf}
the possibility of 
detecting 
solar antineutrinos with the KamLAND experiment.
These antineutrinos are predicted by spin-flavor solutions
to the solar neutrino problem.
As we saw before, the recent evidence from SNO  shows that 
a) the neutrino oscillates, only around 34\% of the 
initial solar neutrinos arrive at the Earth as electron 
neutrinos and
b) the conversion is mainly into active neutrinos,
 however a non e,$\mu-\tau$ component is allowed: 
the fraction of oscillation into  non-$\mu-\tau$ neutrinos 
is found to be 
$\cos^2 \alpha= 0.08^{+0.20}_{-0.40}$.
This residual flux could include sterile neutrinos and/or  
the antineutrinos of the active flavors.

\begin{figure}
\centering
\psfig{file=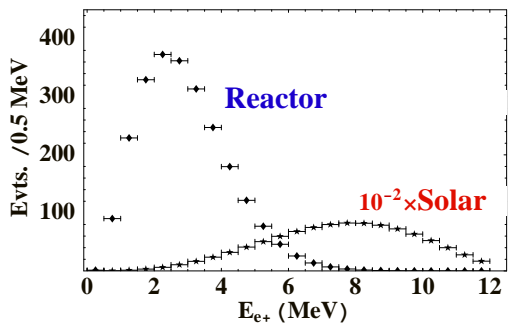,width=7cm}
\label{klafigure} 
\caption{
The KamLAND expected positron spectra 
in absence of oscillations coming from reactor 
antineutrinos (normalized to three years of data taking).
The ``solar'' positron spectrum is obtained assuming the 
shape of the ${}^8 B$ 
neutrino flux and a total normalization
$10^{-2}\times \Phi({}^8 B)$ (that is, an overall 
$\nu_e-\overline{\nu}_e$ 
conversion probability $\overline{P}\sim 1\%$).}
\label{kla:f1}
\end{figure}

The KamLAND experiment is potentially sensitive 
to antineutrinos coming from solar ${}^8$B neutrinos.
Reactor and solar antineutrino signals are shown in 
Fig.(\ref{klafigure}). 
In case of negative results, we find that the 
results of the KamLAND experiment could put 
strict limits on the flux of solar antineutrinos
$\Phi( {}^8 B)< 1.0\times 10^4\ cm^{-2}\ s^{-1}$,
and their appearance probability ($P<0.2-0.15\%$), respectively 
after 1-3 years of operation.
Assuming a concrete model for antineutrino production 
by spin-flavor precession in the convective solar 
magnetic field, this upper bound on the 
appearance probability  implies an upper limit on 
the product of the intrinsic neutrino magnetic moment and
the value of the  field $\sim \mu B< 10^{-21}$ MeV.
For $B\sim 10-100$ kG, 
we would have  $\mu <  10^{-11}-10^{-12}\ \mu_B$.

In the opposite   case, if spin-flavor precession 
is indeed at work even at a minor rate, 
the additional flux of 
antineutrinos could strongly distort the 
signal spectrum seen at KamLAND at energies above 
4 MeV and their contribution should be
taken into account. This is graphically shown in 
Fig.(\ref{kla:f1})

\begin{table}[t]
 \begin{center}
  \scalebox{0.99}{
    \begin{tabular}{lcrccc}
\hline
  E$_{thr}$     & $S_{Sun}$& $S_{Rct}$ & Bckg. &
 P (CL 95)\% & P (CL 99)\% \\
\hline
6 MeV  & 616  & 43 & 70 & 0.22 & 0.23          \\
7 MeV  & 500  & 11 & 65 & 0.19 & 0.20         \\
8 MeV  & 366  & 2 &  60 & 0.21 & 0.23         \\
\hline
      \end{tabular}
}
 \caption{ Expected signals from solar antineutrinos  after 
3 years of data taking. Reactor antineutrino (no oscillation is 
assumed) and other background 
(correlated background) over the same period. The random coincidence 
background is supposed negligble above these energy thresholds. 
Upper limits on the antineutrino oscillation probability.
}
  \label{table1}
 \end{center}
\end{table}

\vspace{0.2cm}
{\bf Acknowledgements}

One of us, V. A., would like to thank all the organizers 
of the ''Third Tropical Workshop on Particle Physics and Cosmology: 
Neutrinos, Branes and Cosmology (Puerto Rico, August 2002)'' 
and in particular J. Nieves, for the kind invitation and for providing 
a pleasant and stimulating atmosphere.  We are glad to thank all the other 
participants of the conference and mainly 
E. Lisi, M. Smy and J. Formaggio for 
very useful discussions.It's a pleasure to thank also 
S.T. Petcov for  enlightening conversations while
this paper was prepared.

\end{document}